\documentclass[twocolumn,amsmath,amssymb,,10pt]{revtex4}

\usepackage{graphicx}
\usepackage{dcolumn}
\usepackage{bm}
\usepackage{amsfonts}
\usepackage[T1]{fontenc}
\usepackage[latin9]{inputenc}

\usepackage{mathrsfs}
\begin{document}
\title{Global and Local Information in Traffic Congestion}
\author{Giovanni Petri$^{1,2}$, Henrik Jeldtoft Jensen$^{2,3}$, John W. Polak$^1$}
\affiliation{
$^1$ Centre for Transport Studies, Department of Civil and Environmental Engineering, Imperial College London, South Kensington Campus, London Sw7 2AZ, UK\\
$^2$ Institute for Mathematical Sciences,   Imperial College London, 53 Princes Gate, London SW7 2PG, UK\\
$^3$  Department of Mathematics, Imperial College London, South Kensington Campus, London Sw7 2AZ, UK}

\begin{abstract}
A generic network flow model of transport (of relevance to information transport as well as physical transport) is studied under two different control protocols. The first involves information concerning the global state of the network, the second only information about nodes' nearest neighbors. The global protocol allows for a larger external drive before jamming sets in, at the price of significant larger flow fluctuations. By triggering jams in neighboring nodes, the jamming perturbation grows as a pulsating core. This feature explains the different results for the two information protocols.   
\end{abstract}
\maketitle

\begin{figure}[!h] 
\centering
\includegraphics[width=0.4\textwidth]{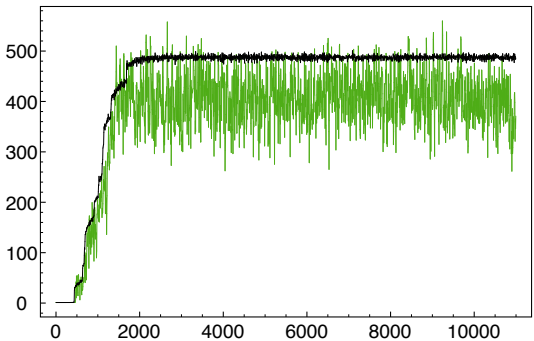}
\caption{Time series plot for global (green) and local (black) models in a RG network with $\beta=0.9$, averaged over realizations.}\label{seed}
\end{figure}

\begin{figure}
\centering
\includegraphics[width=0.4\textwidth]{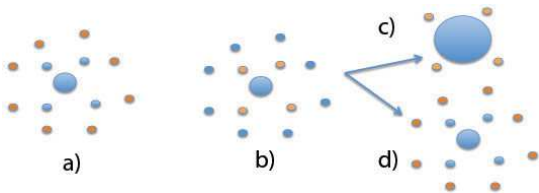}
\caption{Jammed component expansion mechanism.}\label{Core}
\end{figure}

The interplay between information networks and transport networks determine the flow on the latter.  In particular congestion formation, persistence and elimination depend on how information is spread across the transport system. In line with a number of recent studies \cite{DeMartino:2009p777,Scellato:2009p144,Youn:2008p146,Tadic:2004p1036,Sun:2007p2017} we model here these issues from a network theoretical perspective \cite{Barabasi:2002p192,Dorogovtsev:2008p788}. Our aim is to study simple models to obtain qualitative as well as semi-quantitative insight into fundamental aspects of the dynamics of network congestion. 
Transport networks have indeed been object of many studies in recent years. In analogy to equilibrium fluid dynamics, steady-state descriptions \cite{PhysRevE.53.2366} were proposed and studied in depth. Although they produce interesting results about local phenomena \cite{PhysRevE.51.1035,Hewer:2008p799,Mckee:2009p2256,Yuan:2009p2258} they are not able to capture the inherent "intelligent" behaviour involved in route choice. Micro-level agent-based models \cite{763805,Aboratory99losalamos,Daganzo:2006p2361} solve the issue of individual choice, but require an enormous amount of detailed informations about the network (e.g. position and signaling of junctions, traffic lights synchronizing etc.).\\
We propose here a minimal network flow model with unjammed/jammed nodes evolving under two different types of information distribution. In the first case we  let each node  have complete information ({\it global model}). In the second case each node only receive information concerning its neighborhood ({\it local model}). Nodes are thought of as stations or junctions in a traffic network, while links as roads or rail-tracks between two nodes. A node $i$  is characterised by its threshold $T_i =\Omega k_i$, where $k_i$ is the node's degree and $\Omega$ a positive constant, representing the maximum load it  can support before jamming, the load $L_i$ representing the load (people, trains, cars..) present on the node,  and the state  $S_i$, which is set to 1 or 0, respectively when the station is unjammed or jammed. \\
When a node $i$ jams, no incoming flow from neighboring nodes is permitted until $L_i$ becomes smaller then $T_i$ again. This mimicks the behavior of real systems, where a station can block due to too much incoming flow, effectively cutting off its neighbors. Normal operation is restablished after sufficient off-loading. 
The equation for the state of node $i$ is $S_i(t+1) =  \Theta \left( T_i - L_i (t)\right) $, where $\Theta$ is the Heaviside step function, and the outgoing flow $F^{out}_{ij}$ from node $i$  to node $j$ is
\begin{equation}
F^{glob/loc}_{ij}(t) = \frac{min\left(L_i(t) , T_i \right)}{k_i} S_j(t) \left( 1 - J(t)\right) \label{Flusso}
\end{equation}
where $J \in [0,1]$ represents the fraction of jammed stations that a station 'sees' or is informed of and the $min$ function is used to put a superior limit on the outgoing flow from a station. Note that, for growing $J$, the outgoing flow from stations decreases in general. Also, the term $S_j$ on the r.h.s. of (\ref{Flusso}) accounts for the impossibility of sending flow to a jammed station ($F_{ij}=0$ if $S_j=0$).   The idea behind this choice is that a station, if informed that many stations are jammed, will try to gradually reduce its outgoing flux to avoid congesting the system any further.\\
\paragraph{Initial Conditions} We drive the system with two different mechanisms. In the first case, we introduce in the network a total load, $L_{tot} = \beta \sum_{i=0}^N T_i = \beta T_{tot}$, controlled by the parameter $\beta$, that represents the filled fraction of the total network capacity.
The total load is distributed in the network randomly, the only condition being avoiding to jam nodes from the beginning. Operatively, one can accomplish this by assigning to node $i$ a load equal to a random fraction of the node's threshold and then iterating over the nodes until the load introduced in the network reaches the desired value.  The initial conditions are thus:
\begin{eqnarray}
S_i (0)  =   1 \quad \forall i    \nonumber \\
L_i (0)  =  X_i,  \quad  0 < X_i < T_i  \quad \forall  i  \label{InitialConditions1}
\end{eqnarray}
where $X_i$ a random variable $\in [0, T_i ]$, so that $\sum_i L_i = \beta T_{tot}$. The system is then driven through random redistribution of load among nodes. \\
In the second case, we place the entire load on one single node, denoted as the {\it seed} , and study how the system relax without any further drive. So the initial conditions are:
\begin{eqnarray}
S_i (0)  =   1, \quad L_i (0)  =  0  \quad \forall  i\neq seed  \nonumber \\
S_{seed}=0, \quad L_{seed}=L_{tot}  \label{InitialConditions2}
\end{eqnarray}
\paragraph{Dynamics} 
The parameter $J$ in (\ref{Flusso}) represents the fraction of jammed nodes that a station sees in the system. In the local dynamics, each station has information only about its nearest neighbors, thus  $J$ is node-dependent, $J=J^{loc}_i (t)$:
\begin{equation}
J^{loc}_i (t)  =  \frac1k_i \sum_{\{ j \in I_i \} } \left ( 1 - S_j(t) \right)\\
\end{equation}
where the sum is restricted to the neighborhood of node $i$, while in the global dynamics each station has information about the whole network, therefore producing an unique value for $J$ for all the nodes at each time:
\begin{eqnarray}
J^{glob}(t) & = & \frac1N \sum_{m} \left ( 1 - S_m(t) \right)  \, ,
\end{eqnarray}
We note that the completely jammed and unjammed states, $N_J=N$ and $N_J=0$, are absorbing states. Indeed, if the system reaches the state $N_J=N$ we have $J^{loc}_i=J^{glob}=1$ $\forall i$, implying $F_{ij} = 0$ $\forall i,j$. Similarly, if $N_J=0$, consider the load $L_i(t_0)$ on node $i$ at time $t_0$. Since all the neighbors of $i$ are unjammed, the total outgoing flow from $i$ is $\sum_{j\in I_i}F_{ij} = L_i(t_0)$
while the maximum incoming flow is  $\sum_{j\in I_i}F_{ji} = k_i \alpha = T_i$. Thus,
\begin{equation}
L_i(t_0 + 1) = L(t_0) - \sum_{j\in I_i}F_{ij} + \sum_{j\in I_i}F_{ji} \le k_i \Omega = T_i 
\end{equation}
and node $i$ does not jam, since the condition for jamming is $L_i > T_i$. \\
We have seen that starting from $N_J=0$ the system cannot in itself trigger jamming perturbations. However, it is a very unstable state because, in presence of a jammed neighbor, a node keeps some fraction of its previous load on itself. This will move the node closer to the jamming threshold and thus making itself more susceptible to jamming. So if we produce in some way even a single jammed node, we expect  the jamming perturbation to expand through a sort of chain reaction up to a stationary state that depends on the amount of load introduced. This is a first indication that inhomogeneity in the system is the driving force behind jam propagation. 
\paragraph{Simulations}
We performed extensive simulations  on random graph (RG) and scalefree networks(SF) with the number of nodes $N$ varying between $10^2$ and $5 \times 10^4$ under both driving mechanisms. Realisations initialised as in (\ref{InitialConditions1})  were driven by redistributing a fraction of a randomly chosen node's load to another randomly chosen node ($L_i \to  L_i + c L_j , L_j \to L_j (1-c),   c \in (0,1) $). When a jam appeared, the driving was suspended for the duration of the active phase ($N_J\neq0$). Realizations initialized as in Eq.  (\ref{InitialConditions2}) relaxed to their stationary state without active driving.+
\begin{figure}[h]
\includegraphics[width=0.5\textwidth]{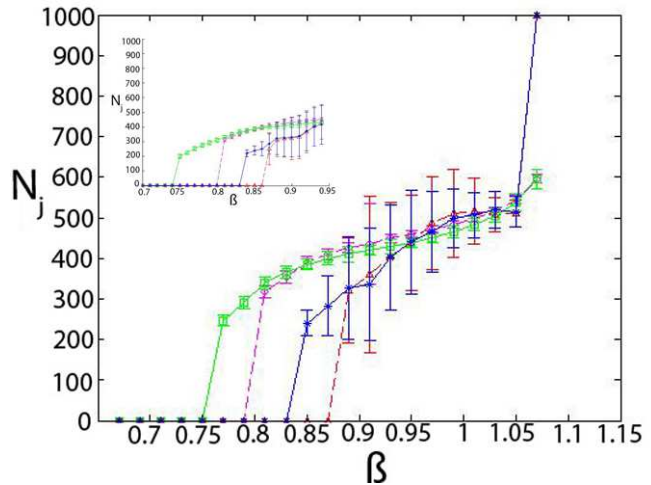}
\caption{$N_J$vs $\beta$ plot for global and local models on SF (full line) and RG (dashed line) networks ($N=1000$) with seeded and uniform (inset) initial conditions. Error bars are standard deviations of fluctuations of $N_J$ around its mean value, averaged over the realizations at a fixed $\beta$. Data shown are obtained over 100 realizations of the RG and SF networks.}\label{phase}
\end{figure}
Figure \ref{phase} shows the characteristic behaviors observed for the asymptotic jammed population $N_J(t\to\infty)$. For $\beta < 0.75$ the system does not show a stationary population of jammed nodes\cite{jamexplanation}. For higher values, we observe $N_J\neq0$ for different values of $\beta_0$ depending on the topology and dynamics, but, in striking contrast to the naive expectation $N_J\to N$,  as $\beta \to 1$ $N_J$ remains in he vicinity of $N/2$ for a broad range of $\beta$ values.  Indeed the applied load is sufficient to jam the entire network.  However the load is trapped in the jammed nodes and being redirected from there. This prevents a uniform redistribution of the load across the network keeping $N_J$ far below its maximal possible value. 
We notice that in the local model the increase in $N_J$ sets in for lower values of $\beta$ than in the global model. Also, the SF networks appear to be more susceptible of developing a stationary jammed population than the RG network. There reason for this is the higher abundance of low degree nodes in SF networks than in RG networks. The low degree nodes are more easily jammed and  provide the core needed for seeding a larger jamming perturbation. Indeed,  $N_J(t)$ under the seeded driving (fig.\ref{seed}) increases through a sequence of jumps and quasi-plateaus, these are the signature of an {\it oscillating core mechanism} (fig.\ref{Core}): the initial jammed seed makes the neighboring nodes more vulnerable to jamming by sending flow toward them, while not accepting flow from them.  This brings the seed's neighbors closer to their threshold. If there is enough load on the seed node, it will eventually jam its own neighborhood (fig. \ref{Core}a); the neighbors will relax onto their unjammed neighbors, but this will make the second neighbors vulnerable or even jam (fig. \ref{Core}b); this mechanism can continue as long as the core has flow to send. The growth of the core can end in two scenarios, a) the seed node's neighbors manage to distribute the load without jamming their own unjammed neighbors and so the load is distributed freely outward (fig. \ref{Core}d), or b) the second neighbors jam and while they relax, they jam the seed's first neighbors again thus creating a bigger, more stable jammed core (fig.\ref{Core}c). The quasi-plateaus in $N_J(t)$ correspond to periods of growth of the vulnerable population, and are followed by sharp jumps, where the jammed core grows by rapidly invading the vulnerable nodes. 
Another interesting feature is that the local model exhibits much narrower fluctuations ($\sigma_l \simeq 30$ for $N=10^3$) than the global model ($\sigma_g\simeq150$). These fluctuations are another signature of the core expansion mechanism: when the systems reaches its stationary $N_J$ value, it keeps fluctuating around the value due to the constant overshooting between jammed and unjammed nodes. The more violent fluctuations observed under the global dynamics are indeed consistent with the core mechanism: the global dynamics, through its global flow suppression, produces smaller "packets" of flow, that allow nodes to stay nearer to their thresholds than under the local dynamics, and therefore produces a larger population of nodes susceptible of jamming, thereby generating bigger fluctuations. For this same reason, the global dynamics develops a stationary jammed population for $\beta$ values larger than for the local dynamics but exhibits also a steeper increase with $\beta$. 
\begin{figure}[!h]
\centering
\includegraphics[width=0.4\textwidth]{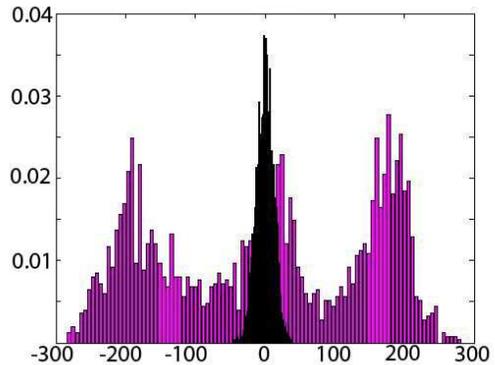}
\caption{An example of fluctuations PDF  for RG for $\beta\simeq0.91$($N=1000$).}\label{SFfluct}
\end{figure}
\paragraph{Fokker-Planck Rate Model}
Our deterministic flow model is computationally heavy to simulate and analytically hard to describe. Although the model is deterministic, the fluctuations due to spatial inhomogeneity suggest it may be possible to understand observed behavior by use of a set of simple transition processes. We separate the population of nodes in to three types:  unjammed (U), vulnerable (V) and jammed (J), with rate equations,
\begin{eqnarray}
U(n,t)+J(m,t)&\stackrel{\alpha}{\rightarrow}&V(n,t)+J(m,t) \\ \label{alpha}
V(n,t)+J(m,t)&\stackrel{\nu}{\rightarrow}&J(n,t)+J(m,t) \\ \label{nu}
J(n,t)+U(m,t)&\stackrel{\gamma}{\rightarrow}&U(n,t)+U(m,t)  \label{gamma}
\end{eqnarray}
Instead of treating individual nodes, we consider degree classes and introduce $U_k$,$V_k$,$J_k$ as the probabilities of choosing randomly a member of a $k$-degree node belonging to one of the three types.
%
%
The rate terms are very intuitive and can be written down directly from the rate equations, considering the contributions from the different degree classes. For example, the contribution of (\ref{nu}) to $J_k$ is of the form $\Gamma_\nu^J(k,t) =  \nu \sum_{k'=1}^{k_{max}} P(V_k(t),J_{k'}(t))$. Later on, we will set $\nu=1$ as definition of vulnerable population.
%
\begin{figure}[!ht]
\centering
\includegraphics[width=0.4\textwidth]{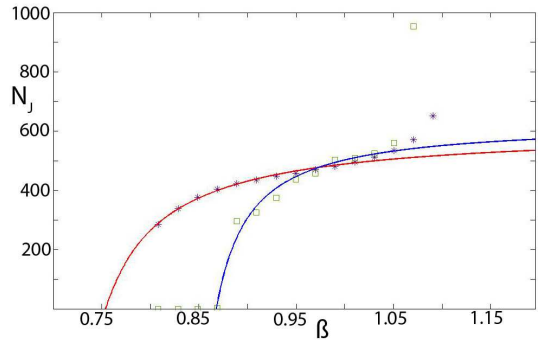}
\caption{SF simulated data ({\it dots}) and fit ({\it line}) obtained using eq. \ref{J} for local (green and blue) and global (violet and red) models.} \label{fit}
\end{figure}
For the moment we neglect degree-degree correlations and write the probabilities $P(a,b)$ as the probability of randomly choosing a node belonging to $a$  and a second node belonging to $b$ times the probability of having a link between the two. For example, $P(V_k,J_{k'})= \frac{V_k}{N} \frac{kk'}{2M}\frac{J_{k'}}{(N-1)}$, in other words the process rate times the number of relevant couples $VJ$. 
If we now substitute the $\Gamma$ terms inside the FP equations for the populations we obtain, for each degree class $k$, the set of equations:
\begin{eqnarray}
J_k(t+1) & = & J_k(t) +  \frac{k}{N(N-1)2M} \times \label{FPJ}\\ 
&& \sum_{k'=1}^{k_{max}} k' \left[ \nu V_{k}(t)J_{k'}(t) - \gamma J_{k}(t)U_{k'}(t)\right] \nonumber \\  
U_k(t+1) & = & U_k(t) +  \frac{k}{N(N-1)2M} \times \label{FPU}\\
&& \sum_{k'=1}^{k_{max}} k' \left[ \gamma J_{k}(t)U_{k'}(t) - \alpha U_{k}(t)J_{k'}(t)\right] \nonumber\\  
V_k(t+1) & = & V_k(t) +  \frac{k}{N(N-1)2M} \times \label{FPV}\\
&& \sum_{k'=1}^{k_{max}} k' \left[ \alpha U_{k}(t)J_{k'}(t) - \nu V_{k}(t)J_{k'}(t)\right] \nonumber
\end{eqnarray}
with $J_k + U_k +V_k = p_k$ and $\sum_k p_k =1$. Performing the sum over $k'$we obtain the average degree for the respective populations. The (\ref{FPJ}-\ref{FPU}) cannot be solved easily. We restrict to the case  $\langle k_J \rangle = \langle k_U \rangle = \langle k_V \rangle$, a strong assumption that is however confirmed by the simulations. So for the stationary state we find:
\begin{equation}
J_k =\frac{\nu}{\gamma}V_k  \quad U_k =\frac{\nu}{\alpha}V_k \quad  V_k =\frac{p_k}{1+ \frac{\nu}{\alpha}+\frac{\nu}{\gamma}}
\end{equation} 
Assuming a simple proportionality between $\beta$ and the "jamming rate" $\alpha \propto (\beta-\beta_0)$ , we can compare the Fokker-Planck prediction 
\begin{equation}
J(\beta)\propto \frac{ (\beta-\beta_0)}{\gamma + (\gamma+1)(\beta-\beta_0)}  \label{J}
\end{equation} 
with the simulated data. Figure \ref{fit} shows this comparison and we can easily see that the Fokker-Planck solution reproduces well the behavior of the simulated data up to $\beta\simeq1$. In particular, the fit correctly identifies the value of $\beta_0$ and reproduces the steep transition from no stationary non-zero $N_J$ to a significant non-zero $N_J$. As expected,  eq. \ref{J} does not predict the second jump in $N_J\to N$: the approximation of uncorrelated degrees and constant rates break down. Indeed, eqs. (\ref{alpha}-\ref{gamma}) are not valid anymore as the system is better described by only two competing population V and J. Moreover, our mean-field-like rate model does not account for random fluctuations that could bring the system in the absorbing $N_J=N$ state. 
\paragraph{In conclusion} we have identified how a jammed core is able to push neighboring nodes to their thresholds and eventually invade them. We also demonstrated that the network topology and the different information regimes (local and global) play a significant role only at the onset of jams (by shifting threshold load level $\beta$), but do not significantly influence  the rest of the evolution, which is qualitatively well described by a model of 3 competing populations, mimicking the unjammed (U), vulnerable (V) and jammed (J) populations of the flow model. 

Of relevance to traffic control we have shown that under global dynamics the jam picks up at a larger driving load than when only local information is used. However, in our model global traffic management  exhibits much larger fluctuations and a steeper increase in number of jammed node. Indeed, for real systems, our model suggests that under heavy loads it is better to let the system evolve under local information rather than attempting   global management, since jams sweep a much larger part of the system in the latter case. Moreover, since congested regions enhance the jamming perturbation by making their neighbors more susceptible to jamming, it is better to reduce the number of links connecting the congested regions to non-congested ones (in some sense, quarantine them) rather than letting them communicate with the neighboring regions and thereby risk the triggering of new jams. 

\bibliographystyle{phcpc}

\begin{thebibliography}{10}
\bibitem{jamexplanation}
A transient jammed population is indeed observed but it decays into $N_J=0$. .
\end{thebibliography}

\end{document}